\documentclass[12pt]{iopart}
\usepackage[utf8]{inputenc}
\usepackage{graphicx}
\usepackage{siunitx}
\usepackage{xcolor}
\usepackage{subfig}
\usepackage{tabularx}
\usepackage{iopams}
\expandafter\let\csname equation*\endcsname\relax
\expandafter\let\csname endequation*\endcsname\relax
\usepackage{amsmath}

\begin{document}

\title{Thermal Noise in Electro-Optic Devices at Cryogenic Temperatures}

\author{Sonia Mobassem$^{1,2}$, Nicholas J. Lambert$^{1,2}$,  
Alfredo Rueda$^{1,2,3}$, Johannes M. Fink$^{3}$, Gerd Leuchs$^{1,2,4,5}$,\\ and Harald G. L. Schwefel$^{1,2}$} 

\address{\small$^1$The Dodd-Walls Centre for Photonic and Quantum Technologies, New Zealand\\
	$^2$Department of Physics, University of Otago, New Zealand\\
	$^3$Institute of Science and Technology Austria, Klosterneuburg, Austria\\
	$^4$Max Planck Institute for the Science of Light, Erlangen, Germany\\
	$^5$Institute of Applied Physics of the Russian Academy of Sciences, Nizhny Novgorod, Russia}
\ead{nicholas.lambert@otago.ac.nz}  
\vspace{10pt}
\begin{indented}
\item[]\today
\end{indented}

\begin{abstract}
The quantum bits (qubit) on which superconducting quantum computers are based have energy scales corresponding to photons with \SI{}{GHz} frequencies. The energy of photons in the gigahertz domain is too low to allow transmission through the noisy room-temperature environment, where the signal would be lost in thermal noise. Optical photons, on the other hand, have much higher energies, and signals can be detected using highly efficient single-photon detectors. Transduction from microwave to optical frequencies is therefore a potential enabling technology for quantum devices. However, in such a device the optical pump can be a source of thermal noise and thus degrade the fidelity; the similarity of input microwave state to the output optical state. In order to investigate the magnitude of this effect we model the sub-Kelvin thermal behavior of an electro-optic transducer based on a lithium niobate whispering gallery mode resonator. We find that there is an optimum power level for a continuous pump, whilst pulsed operation of the pump increases the fidelity of the conversion.

\end{abstract}

%
%
%
%
%

\section{Introduction}

Quantum computers based on superconducting qubits~\cite{devoret_superconducting_2013} have seen enormous progress in recent years. Quantum coherence is preserved as a consequence of reduced dissipation and the absence of thermal photons, but this typically requires cooling to below \SI{100}{\milli\kelvin} for qubits with characteristic frequencies of $\sim\SI{10}{\giga\hertz}$~\cite{devoret_implementing_2004}. Communication between spatially separated qubits therefore presents a particular difficulty. Outside the cryogenic environment, at room temperature, the microwave frequency photons carrying the quantum information are  completely swamped by thermal photons; however, visible or near infrared photons are of much higher energy and are known to be able to carry quantum information over long distances at room temperature. Therefore, a future quantum network \cite{kimble_quantum_2008,wehner_quantum_2018} will need a method of up- and down-converting   microwave photons to near infrared or visible frequency photons, to enable coherent communication between different quantum systems~\cite{divincenzo_physical_2000, simon_towards_2017, huang_observation_1992}.

A number of approaches to microwave up-conversion have been explored \cite{lambert_coherent_2020, clerk_hybrid_2020, kurizki_quantum_2015,lauk_perspectives_2020}, using the non-linearities offered by magneto-optic materials~\cite{williamson_magneto-optic_2014, fernandez-gonzalvo_coherent_2015,everts_microwave_2019,haigh_triple-resonant_2016,hisatomi_bidirectional_2016}, cold atom clouds~\cite{hafezi_atomic_2012,han_coherent_2018,vogt_efficient_2019,zibrov_four-wave_2002}, quantum dots~\cite{rakher_quantum_2010, singh_quantum_2019} and opto-mechanical devices~\cite{andrews_bidirectional_2014,bagci_optical_2014,stannigel_optomechanical_2010,higginbotham_harnessing_2018}. The effect of this non-linearity is often increased by resonant enhancement of one or more of the input field, output field and optical pump. Here we focus on an electro-optic architecture~\cite{strekalov_microwave_2009,tsang_cavity_2010,javerzac-galy_-chip_2016,soltani_efficient_2017,zhang_electro-optic_2018,fan_superconducting_2018,witmer_silicon-organic_2020}, in which the required non-linearity is provided by the electro-optic tensor of lithium niobate (LiNbO$_3$)~\cite{rueda_efficient_2016,hease_cavity_2020}.  The input field mode is defined by a metallic microwave cavity, and the output and pump by whispering gallery modes (WGMs)~\cite{strekalov_nonlinear_2016} confined to the perimeter of a LiNbO$_3$ disc.

The optical pump is a feature common to all up-conversion techniques, providing both the energy for the up-converted photon, and a reference frequency. The efficiency of the up-conversion process is typically improved by increasing the power in the pump, which may be as high as \SI{1.48}{\milli\watt}~\cite{hease_cavity_2020}, or even  \SI{6.3}{\milli\watt}~\cite{fan_superconducting_2018} in an electro-optic realization. 
However, a significant fraction of the optical pump is dissipated in the device, and this can conflict with the cryogenic requirements for the environment in which superconducting qubits must be hosted. Dilution fridges, which are the usual method of attaining millikelvin temperatures, have cooling powers not significantly exceeding \SI{1}{\milli\watt}~\cite{uhlig_dry_2008, noauthor_xld_nodate} at \SI{100}{\milli\kelvin} with around \SI{400}{\micro\watt} being more typical. Higher dissipated powers result in a dynamical thermal equilibrium with elevated temperatures. A balance exists, therefore, between choosing a pump power large enough for the up-conversion efficiency to be useful, but not so large that heating produces a high enough thermal photon population that the quality of the quantum state transfer is compromised.  

The quality of the state transfer is characterized by the fidelity, which is defined as the overlap between the input and output quantum states; a fidelity of 1 corresponds to equal states (perfect conversion), and 0 to orthogonal states. Long distance quantum telecommunication~\cite{braunstein_quantum_2012}, remote quantum state preparation~\cite{laurat_conditional_2003,pogorzalek_secure_2019} and quantum state coherent computing~\cite{neergaard-nielsen_optical_2010} are often based on continuous variable (CV) states, and therefore, the fidelity between two such states lies on a continuum between 0 and 1. Examples include states of the type $|\psi_c\rangle=(|\alpha\rangle\pm|-\!\alpha\rangle)/\sqrt{2}$ (so-called `cat' states), which are commonly used as CV-qubits~\cite{neergaard-nielsen_optical_2010}, and squeezed states such as $|\psi_s\rangle=|\alpha, r\rangle$, which are useful in telecommunication schemes~\cite{braunstein_quantum_2012, weedbrook_gaussian_2012}. 

\begin{figure}
  \centering
  \includegraphics[width=1\textwidth]{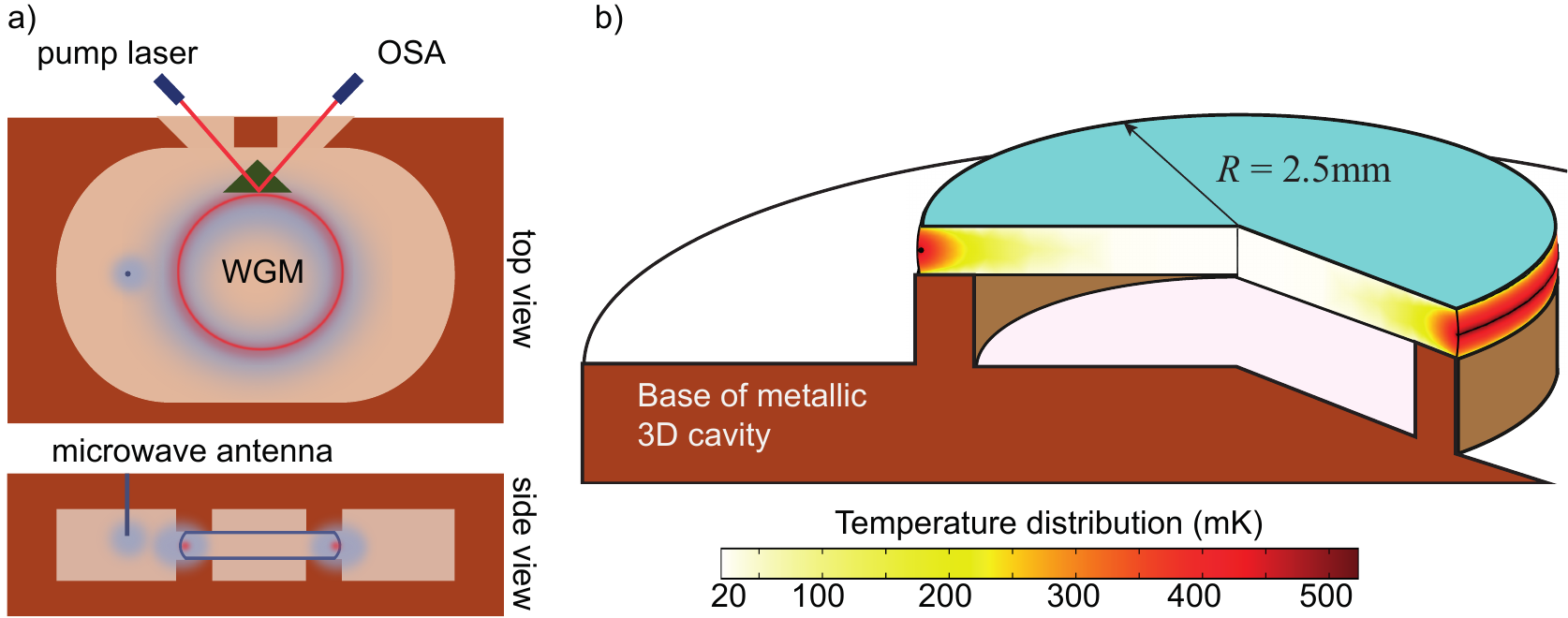}
  \caption{a) Schematic of the microwave cavity with an embedded dielectric whispering gallery mode (WGM) resonator. Light is coupled through a silicon prism and the microwave tone is inserted through an antenna. The WGM is clamped between metal rings to confine the microwave field to the optical mode volume. b) Numerical finite element simulation of the temperature distribution within the dielectric WGM assuming a critically coupled optical pump with $P_o=\SI{1}{\milli\watt}$. The initial temperature of the device was $T_\mathrm{B}=\SI{20}{\milli\K}$. The optical mode is indicated by the tiny dot close to the rim. Only one half of the metallic cavity is shown.}\label{fig:Setup}
\end{figure}

Up-conversion of thermal noise to the optical channel results in a reduced fidelity of state transfer. However, so does a low conversion efficiency.
In order to investigate this compromise further, we carried out thermal simulations of an efficient electro-optic modulator consisting of an optical resonator coupled to a microwave cavity~\cite{rueda_efficient_2016, rueda_resonant_2019,hease_cavity_2020}. The optical resonator is a WGM resonator  made of monocrystaline LiNbO$_3$, into which light is coupled using a silicon prism (see Fig.~\ref{fig:Setup}a). The convex geometry of the WGM resonator and transparency of LiNbO$_3$ confine light in the rim of the resonator via total internal reflection  (see Fig.~\ref{fig:Setup}b).  The confinement of the optical mode in a very small mode volume increases the intensity of the optical field and enhances the nonlinear interaction of light in the LiNbO$_3$. The WGM resonator is embedded in a 3D copper microwave cavity, and the geometry of both resonators maximize the overlap between microwave field and optical mode.

In order to study the effect of absorption-induced heating on our electro-optic device we numerically model heat transport in a representative geometry comprising a LiNbO$_3$ WGM resonator with major radius $R = \SI{2.5}{\mm}$ and side curvature $R_c = \SI{1.45}{\mm}$, embedded in a copper cavity (Fig.~\ref{fig:Setup}). The heat transfer partial differential equations are solved using COMSOL Multiphysics software, which implements a finite element method combined with a stiff ordinary differential equation solver~\cite{noauthor_comsol_nodate,edsberg_introduction_2015}.
\begin{table}[b!]
\centering
\caption{Extrapolated thermal characteristics of LiNbO$_3$ and copper with residual resistance ratio $\text{RRR}=100$ (thermal conductivity) and $\text{RRR}=30$ (heat capacity) at temperatures below 1 Kelvin, assuming the functional behaviour given in Ref.~\cite{krinner_engineering_2019}. }\label{Table:Thermalcharacteristis}
\begin{tabular}{c c c c}
\hline
  & Thermal conductivity (\si{\watt\per\meter\per\kelvin})  & Heat capacity  (\si{\joule\per\kilogram\per\kelvin}) & Ref \\
  \hline
  LiNbO$_3$ & $4\cdot T^3$ & $2.705\times10^{-4}\cdot T^3$&\cite{perez-enciso_thermal_1998}
 \\
  Cu & $500\cdot T$ & $0.01\cdot T$  &\cite{duthil_material_2015} \\
\hline\\
\end{tabular}
\end{table}

Experimental values for the thermal conductivity and heat capacity of LiNbO$_3$ are only available down to \SI{4}{\kelvin} and for copper to around \SI{1}{\kelvin}  \cite{perez-enciso_thermal_1998,duthil_material_2015,krinner_engineering_2019,woodcraft_recommended_2005}. The thermal conductivity and specific heat capacity of dielectrics such as LiNbO$_3$ have a $T^3$ dependence at cryogenic temperature. On the other hand, at very low temperatures the thermal conductivity of copper is proportional to the temperature. To estimate thermal conductivity and specific heat capacity of the materials, we make appropriate downwards extrapolations from data above \SI{4}{\K}, as detailed in Table \ref{Table:Thermalcharacteristis}. We neglect the thermal contact resistance at the LiNbO$_3$-Cu interface, due to the smoothness of the LiNbO$_3$ and the softness of the copper generally giving a good interface, and the low thermal conductivity of the dielectric dominating heat transport (see Table~\ref{Table:Thermalcharacteristis}).\\

\begin{figure}[t]
	\centering
		\includegraphics[width=1\textwidth]{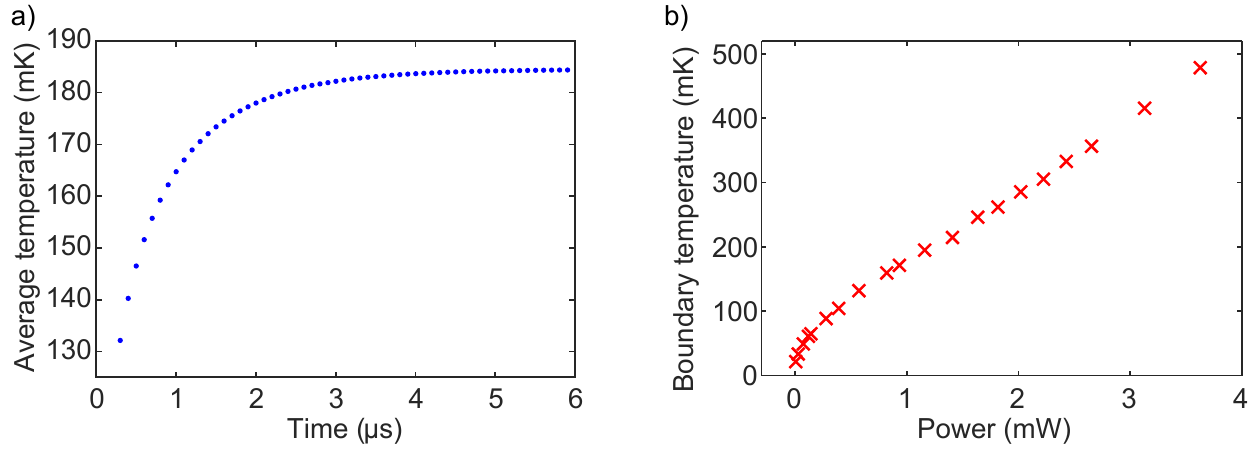}
		\caption{(a) Time evolution of average temperature of the microwave mode volume in a LiNbO$_3$ resonator with \SI{200}{\micro\meter} thickness and critically coupled optical pump with $P_o=\SI{1}{\milli\watt}$. Initially the cavity is at $T_\mathrm{B}=\SI{20}{\milli\K}$ and heats within a timescale of $\SI{3}{\micro\second}$ to $T_\mathrm{max}=\SI{184}{\milli\kelvin}$.  (b) Dilution fridge mixing chamber temperature $(T_\mathrm{B})$ as a function of dissipated power.   }\label{fig:TimeEvolution}

\end{figure}

In the heat transfer simulation of our microwave to optical converter, the optical pump is modelled as a heat source occupying the same volume as the optical mode. The dissipated power follows an exponential approach to the pump power $P_o$ as $P=P_o(1-\exp(- t/\tau))$, where $\tau$ is the rise time of optical mode (for our optical resonator $\tau\sim \SI{1}{\micro\second}$, corresponding to a quality factor of $Q\sim10^8$). The thermal boundary conditions are given by the temperature of the mixing chamber plate of the fridge. As the fridge cooling power at $T\sim\SI{100}{\milli\kelvin}$ is comparable to the heat load introduced by the optical pump, the boundary temperature $T_\mathrm{B}$ for CW operation is also a function of the power dissipated in our device.  We determine typical mixing chamber temperatures for different heat loads by putting a known load on a resistive heater mounted at the mixing chamber of our BlueFors LD250, and then waiting for the temperature to stabilize (see Fig.~\ref{fig:TimeEvolution}(b)). The precise details of this dependency will vary between cryogenic technologies and environments.

Conversely, for pulsed operation of the optical pump, the average heat load will be decreased by the duty cycle of the pump; for example, if the pump is on for only 1\% of the operation cycle, then the heat load will be 1\% of the pulsed power. The temperature of the cryogenic environment will then remain close to the base temperature of the fridge. To study this mode of operation, we fix our thermal boundary at $ T_\mathrm{B}=\SI{20}{\milli\kelvin}$.

We expect the optical rise time, and therefore heating, of the device to be orders of magnitude slower than the diffusion of the heat within the dielectric. By considering the lowest order component of solutions to the heat diffusion equation, we find $t\sim \rho Cd^{2}/K$, with heat capacity $C$, thermal conductivity $K$ and density $\rho$. For LiNbO$_3$ with a thickness of $2d=\SI{200}{\micro\meter}$, the characteristic time for the heat to be redistributed is $t=\SI{3}{\nano\second}$.  

The evolution of the temperature in the LiNbO$_3$, considering the full geometry, is calculated as a function of time, with the LiNbO$_3$ temperature at $t=0$ set to the same temperature as the mixing chamber $ T_\mathrm{B}=\SI{20}{\milli\kelvin}$. In Fig.~\ref{fig:TimeEvolution}(a) we plot the average temperature $T_\mathrm{av}$ of LiNbO$_3$ integrated over the microwave mode volume within the dielectric as a function of time for a WGM resonator thickness of \SI{200}{\micro\meter} and a pump power of \SI{1}{\milli\watt}. The heating occurs on a timescales of $\sim \SI{3}{\micro\second}$ for the parameters of our cavity. We compare this calculation for the case where the optical power immediately reaches its maximum within the optical mode volume. Here the temperature saturates after only $\sim\SI{6}{\nano\second}$, corresponding well to the analytical toy model above.    

 \begin{figure}[t]
	\centering
	\includegraphics[width=1\textwidth,clip]{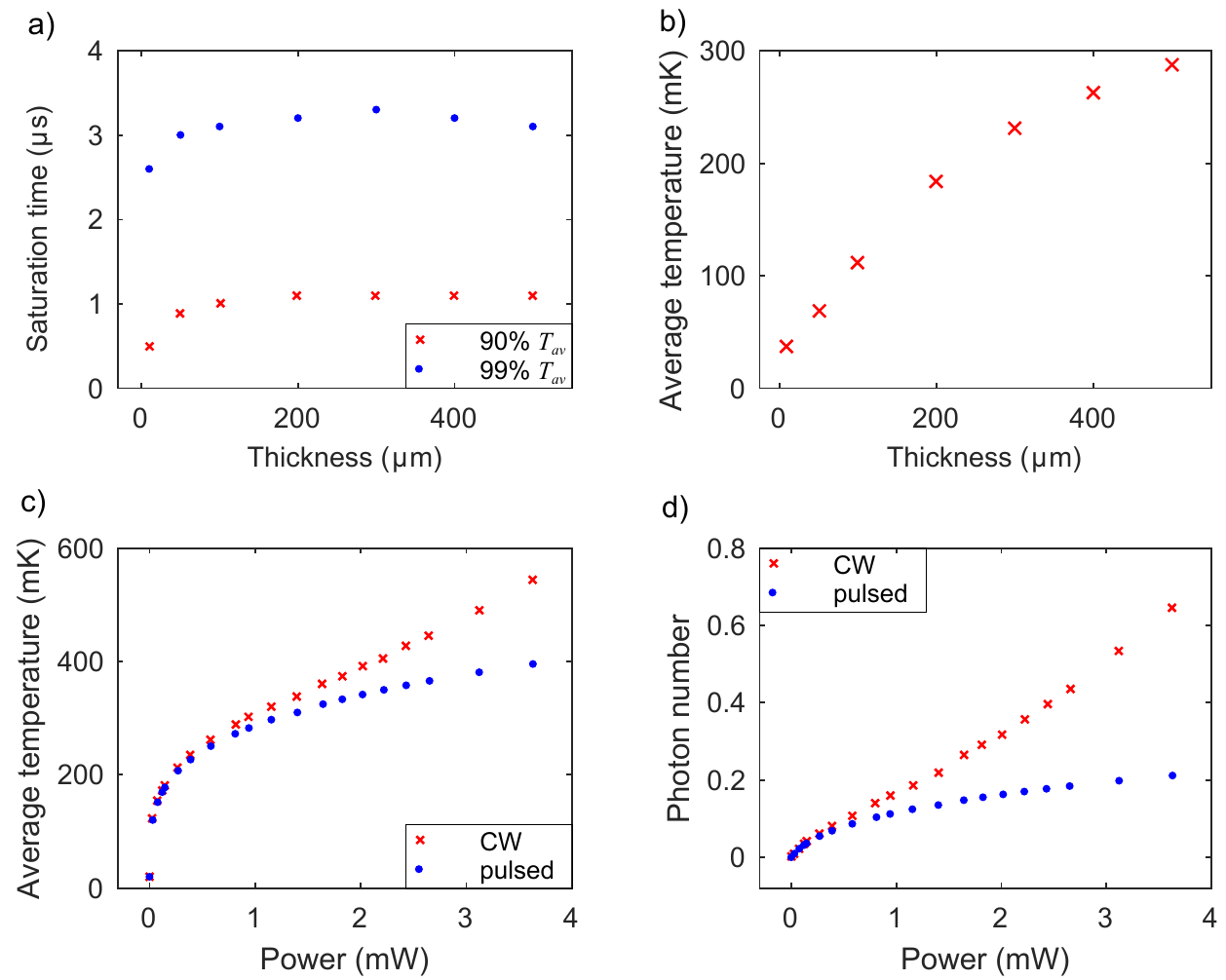}
	\caption{Numerical simulation of the optical heating of the LiNbO$_3$ resonator. (a) Time to reach 90\% (red) and 99\% (blue) of the maximum temperature integrated over the microwave mode volume within the LiNbO$_3$ resonator versus thicknesses of the resonator for $P_0=\SI{1}{\milli\watt}$. (b) Average equilibrium temperature of the mode volume within the LiNbO$_3$ resonator for different thicknesses for $P_0=\SI{1}{\milli\watt}$ at $t=\SI{0.1}{\milli\second}$. (c) Average equilibrium temperature of microwave mode volume as a function of optical pump power for CW (red) and pulsed (blue) operation for \SI{500}{\micro\meter} thickness at $t=\SI{0.1}{\milli\second}$.  (d) Thermal photon occupancy as a function of optical pump power for CW (red) and pulsed (blue) operation for \SI{500}{\micro\meter} thickness at $t=\SI{0.1}{\milli\second}$.}
	\label{fig:fourplots}
\end{figure}

One of the parameters affecting the time evolution of temperature is the thickness of the LiNbO$_3$ resonator, which is often reduced in order to increase the strength of the electric field across it. In Figs.~\ref{fig:fourplots}(a) and \ref{fig:fourplots}(b) we show the saturation time and the asymptotic average temperature as a function of thickness. Since LiNbO$_3$ has a very low thermal conductivity compared to copper, an increase in thickness corresponds to an increase in average temperature, as well as an increase in the time taken to reach that temperature.

We now fix the thickness of the resonator at \SI{500}{\micro\meter}, and study the equilibrium thermal occupancy of the microwave cavity as a function of pump power. For a mode of temperature $T$ and angular frequency $\omega$, the thermal photon number occupancy is
\begin{equation}
n=\frac{1}{\exp\Bigl(\frac{\hbar \omega}{k_BT}\Bigr)-1}.\label{eq:photonNumber}
\end{equation}
The total thermal photon number of the mode is given by 
\begin{equation}
n_{\mathrm{th}}=\frac{\kappa_{\mathrm{i}}}{\kappa}n_{\mathrm{th,i}}+\sum_{j}\frac{\kappa_{\mathrm{e,j}}}{\kappa}n_{\mathrm{th,j}}.\label{eq:TotalPhotonNumber}
\end{equation}
Here $n_{\mathrm{th,j}}$ is the thermal occupancy of the $j$th port of the cavity. $\kappa=\kappa_{\mathrm{i}}+\sum_{j}\kappa_{\mathrm{e,j}}$ is the total intensity loss rate of the cavity and is the sum of the internal loss rate, $\kappa_{\mathrm{i}}$, and the loss rate via the $j$th port  $\kappa_{\mathrm{e,j}}$~\cite{lambert_coherent_2020}.
$n_{\mathrm{th,i}}$ is the thermal occupancy of the mode due to internal fluctuations. This is directly related to internal losses by the fluctuation-dissipation theorem~\cite{gardiner_quantum_2004}. Because of the large loss tangent of LiNbO$_3$, dielectric losses are the dominant contribution to $\kappa_i$. The microwave electric field within the LiNbO$_3$ is uniform, and so we use the spatial average of the temperature within the microwave mode volume inside the dielectric (Fig.~\ref{fig:fourplots}(c)), as calculated for the asymptotic temperature distribution.

We calculate thermal photon numbers for the microwave mode of our single port cavity with angular frequency $\Omega=2\pi\times\SI{10}{GHz}$ as a function of pump power, while fixing the coupling rate of the port to critical coupling ($\kappa=\kappa_{\mathrm{i}}+\kappa_{\mathrm{e,1}}$,  $\kappa_{\mathrm{i}}= \kappa_{\mathrm{e,1}}$).  In Fig.~\ref{fig:fourplots}(d) we present data for two possibilities, one in which the port temperature rises with increasing pump power due to the elevated mixing chamber temperature (CW operation of the optical pump) and one in which the port temperature remains constant at \SI{20}{\milli\kelvin} (pulsed operation of the pump).

\begin{figure*}[t]
	\centering
		\includegraphics[width=0.95\textwidth]{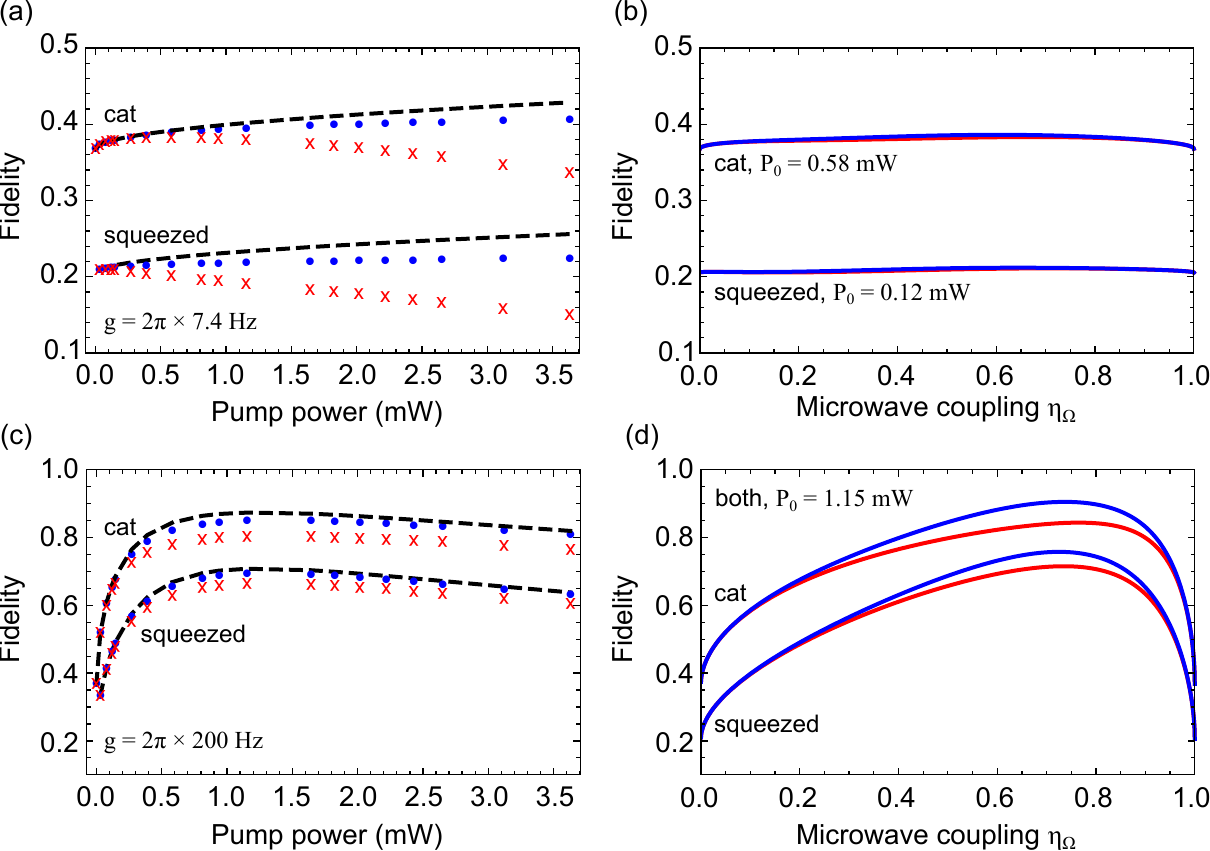}
	\caption{Dependence of fidelity on optical power and microwave coupling. (a) and (c) Fidelity versus optical pump power for cat (top) and squeezed (bottom) states for critical microwave coupling and overcoupled optical fields ($\kappa_{o,e}=4\kappa_{o,i}$). Panel (a) is for $g = 2\pi\times\SI{7.4}{\hertz}$, and panel (c) for an optimised configuration with $g = 2\pi\times\SI{200}{\hertz}$. Results for pulsed (blue circles) and CW (red crosses) operation are shown, and dashed lines correspond to the system with no heating (where everything remains at $\SI{20}{\milli\kelvin}$). (b) and (d) fidelity versus microwave coupling for optimal pump powers in panels (a) and (c). Maxima in fidelity are observed in the overcoupled regime.}
\label{fig:fidelity}
\end{figure*} 
We calculate the effect of the simulated thermal noise on the fidelity of converted states using input–output theory from Ref.~\cite{savchenkov_tunable_2003} for feasible experimental parameters~\cite{rueda_efficient_2016}. 
We use the multi-photon cooperativity $C=\frac{4n_pg^2}{\kappa_o\kappa_\Omega}$, where  $\kappa_{\mathrm{o(\Omega)}}=\kappa_{\mathrm{i,o(\Omega)}}+\kappa_{\mathrm{e,o(\Omega)}}$ is the total loss rate of the optical (microwave) mode, with $\kappa_{\mathrm{i,o(\Omega)}}$ and $\kappa_{\mathrm{e,o(\Omega)}}$ being the intrinsic and the extrinsic damping rates of the optical (microwave) modes, respectively. We calculate the fidelity of the system while the optical resonator is
critically coupled, $\kappa_{\mathrm{i,o}}=\kappa_{\mathrm{e,o}}=2\pi\times\SI{0.7}{\mega\hertz}$, and the
microwave resonator is undercoupled, $\kappa_{\mathrm{e,\Omega}}=2\pi\times\SI{7.2}{\mega\hertz}$ and $\kappa_{\mathrm{i,\Omega}}=2\pi\times\SI{32.4}{\mega\hertz}$. $n_p=\frac{4\eta_o}{\kappa_o}\frac{P_o}{\hbar\omega_p}$ is the photon number corresponding to optical pump power $P_o$, $g$ is the coupling strength and for our system $g=2\pi\times\SI{7.4}{\hertz}$. Here, $\eta_{\mathrm{\Omega(o)}}=\frac{\kappa_{\mathrm{e,\Omega(o)}}}{\kappa_{\mathrm{\Omega(o)}}}$ is the waveguide-cavity coupling.

 The fidelity of the transferred coherent squeezed state $|\alpha,r\rangle$ is given by~\cite{rueda_electro-optic_2019}
\begin{equation}\label{fidelity}
    F^{\mathrm{G}}_{\mathrm{tr}}(\alpha,r,C)=\frac{\exp\Big(-2|\alpha|^2(\epsilon_3-1)^2\big(\frac{\cos(\phi_\alpha)}{V_-}+\frac{\sin(\phi_\alpha)}{V_+}\big)\Big)}
    {\sqrt{\frac{\epsilon_2}{2}(1-\epsilon^4_3)+\epsilon^4_3\Big(1+\frac{\bar{n}_\Omega(\epsilon_2+\epsilon_2\bar{\epsilon}^2_3-2+\frac{\bar{n}_\Omega}{C\eta_o})}{C\eta_o}\Big)}},
\end{equation}
 where $ V_\pm=1+ {\epsilon}^2_3(e^{\pm2r}-1+\frac{2\bar{n}_\Omega}{\eta_oC})$,  $ \epsilon_2=1+\cosh(2r)$,  $\epsilon_3=\frac{\sqrt{4\eta_o\eta_\Omega}}{1+C}$, $|\alpha|$ is field amplitude, $\phi_\alpha$ is the phase of the field, and $\bar{n}_\Omega$ is the equilibrium mean thermal photon numbers of the microwave field.
 The fidelity of the transferred cat state $|1\rangle-|-\!1\rangle$ is given by~\cite{rueda_electro-optic_2019}
\begin{multline}
    F^{\mathrm{cat}}_{\mathrm{tr}}=\frac{1}{\epsilon_4(1+\epsilon_5)}\Bigg[
    \Big(1+e^{\mathrm{\frac{8\alpha^2\epsilon_3}{1+\epsilon_5}}}\Big)e^{\mathrm{-\frac{2\alpha^2(1+\epsilon^2_3)^2}{1+\epsilon_5}}}\\+ 2\cos(\phi)\left(e^{\mathrm{-\frac{2\alpha^2(\epsilon^2_3 +\epsilon_5)}{1+\epsilon_5}}}
    +e^{\mathrm{-\frac{2\alpha^2(1+\epsilon^2_3\epsilon_5)}{1+\epsilon_5}}}\right)\\+\cos(2\phi)e^{\mathrm{-\frac{2\alpha^2(\epsilon_3+\epsilon_5)^2}{\epsilon_5(1+\epsilon_3)}}}+e^{\mathrm{-\frac{2\alpha^2(\epsilon_3-\epsilon_5)^2}{\epsilon_5(1+\epsilon_5)}}}\Bigg],
    \end{multline}
where $\epsilon_4=(1+\cos(\phi)e^{\mathrm{-2\alpha^2}})(1+\cos(\phi)e^{\mathrm{-2\alpha^2\epsilon_3^2}})$ and $\epsilon_5=1+\frac{8\eta_\Gamma\bar{n}_\Omega}{(1+c)^2}$.

The fidelity for states with  cooperativity $C<1$~\cite{rueda_efficient_2016}, is shown vs.\ the optical pump power $P_o$ in Fig.~\ref{fig:fidelity}(a), where the dashed lines show the noiseless case of a cat state $|1\rangle-|-\!1\rangle$ (top) and squeezed state $|\psi\rangle=|1, 0.5\rangle $ (bottom). The red markers indicate CW pump operation and the blue crosses pulsed operation. The maximum fidelity for CW operation is achieved at $P_o\approx \SI{0.58}{\milli\watt}$ for the cat state and $P_o\approx \SI{0.12}{\milli\watt}$ for squeezed state. Fidelities for pulsed operation are shown in blue. These do not exhibit a peak, but still diverge from the fidelities for noiseless operation as the power increases. At $P_o=0$ the fidelity is non-zero due to the state's overlap with the vacuum.
 
The added noise due to dielectric heating causes a degradation of the fidelity of around 4.6\% (cat state) and 4.1\% (squeezed state) from the noiseless case. This problem can be reduced by over-coupling the microwave system; in this case the waveguide temperature, which is always less than the mode temperature, increases the fidelity by acting to cool the mode~\cite{hease_cavity_2020,santamaria-botello_sensitivity_2018,xu_radiative_2020}. In Fig.~\ref{fig:fidelity}(b), we fix the optical pump power to the maximum fidelity of Fig.~\ref{fig:fidelity}(a) with $P_o\approx \SI{0.58}{\milli\watt}$ for the cat state and $P_o\approx \SI{0.12}{\milli\watt}$ for the squeezed state, and vary the microwave coupling from under-coupling ($\eta_\Omega=0$) to over-coupling ($\eta_\Omega=1$). For each state the red curve includes heating of the coupling waveguide (CW operation) and the blue curve describes pulsed operation which leaves the waveguide at the base temperature. We find optimal couplings for pulsed operation of $\eta_\Omega = 0.60$ (cat state) and $\eta_\Omega = 0.66$ (squeezed state), and for CW operation of $\eta_\Omega = 0.61$ (cat state) and $\eta_\Omega = 0.66$.

We now study the effects of thermal noise on an optimised transducer. The principle route to higher efficiencies in electro-optic up-convertors is by increasing the co-operativity $g$. In Fig.~\ref{fig:fidelity}(c) we plot fidelity as a function of pump power for a device with $g = 2\pi\times\SI{200}{\hertz}$. We find that there still exists optimal pump powers. Furthermore, in Fig.~\ref{fig:fidelity}(d) the maximum fidelity is found to be at higher microwave overcouplings, as the cooling effect of the microwave port becomes more pronounced.

Finally, we note that in our calculations, the rise in the temperature of the copper cavity is negligible due to the relatively large heat capacity and thermal conductivity of copper, even at cryogenic temperatures. This may not be the case for superconducting cavities, in which the heat capacity is exponentially suppressed below the superconducting transition temperature. Furthermore, stray photons in a superconducting cavity may lead to a significant non-equilibrium quasiparticle population. For superconducting materials such as aluminium, an appreciable rise in the cavity material temperature may occur on longer timescales~\cite{hease_cavity_2020,mckenna_cryogenic_2020}. Further analysis of this is beyond the scope of the present work.
 
In conclusion, we have demonstrated that the fidelity of quantum state transfer in electro-optic microwave-to-optical transducers is significantly affected by heating due to the absorption of the optical pump. In particular, the choice of pump power must take into account the competition between increased efficiency and increased heating that follow increased optical power. We find that, for some parameter regimes, there is an optimal power which maximises state transfer fidelity. Furthermore, there is also an optimal coupling to the microwave input waveguide which is significantly more than critical coupling. Whilst our calculations have used an electro-optic structure as an archetype, the universality of the need for an optical pump in quantum transducers means that our conclusions can be extended to other platforms. Recently optomechanical micro-devices have been demonstrated with efficient microwave-optical photon conversion~\cite{arnold_converting_2020,han_cavity_2020}. However, they too suffer  from large thermal noise even at millikelvin temperatures when pump power is increased over a certain limit ($\sim\SI{625}{\pico\watt}$~\cite{arnold_converting_2020}, $\sim\SI{40}{\milli\watt}$~\cite{han_cavity_2020}), showing the universal applicability of our analysis.

\section*{Acknowledgements}
N.J.L. is supported by the MBIE Endeavour Fund (UOOX1805) and G.L. is  by the Julius von Haast Fellowship of New Zealand. S.M. acknowledges stimulating discussions with T.M. Jensen.

\section*{References}

\bibliographystyle{unsrt}
\bibliography{reference}

\end{document}